# Electric field tunable coupling strength and quantum metric hot spots in a moiré flatband superconductor

Le Liu[1,2,#], Yu Hong[1,2,#], Chengping Zhang[3,#], Jundong Zhu[1,2], Jingwei Dong[1,2], Kenji Watanabe[4], Takashi Taniguchi[5], Luojun Du[1,2], Dongxia Shi[1,2,6], Kam Tuen Law[3,*], Guangyu Zhang[1,2,6,*], Wei Yang[1,2,6*]

[1] *Beijing National Laboratory for Condensed Matter Physics and Institute of Physics, Chinese Academy of Sciences, Beijing 100190, China*
[2] *School of Physical Sciences, University of Chinese Academy of Sciences, Beijing, 100190, China*
[3] *Department of Physics, Hong Kong University of Science and Technology, Clear Water Bay, Hong Kong, China*
[4] *Research Center for Electronic and Optical Materials, National Institute for Materials Science, 1-1 Namiki, Tsukuba 305-0044, Japan*
[5] *Research Center for Materials Nanoarchitectonics, National Institute for Materials Science, 1-1 Namiki, Tsukuba 305-0044, Japan*
[6] *Songshan Lake Materials Laboratory, Dongguan 523808, China*

\# These authors contribute equally
\* Corresponding authors. Email: wei.yang@iphy.ac.cn; phlaw@ust.hk; gyzhang@iphy.ac.cn

**ABSTRACT**

Superconductivity in flatband systems has attracted tremendous attention in condensed matter physics. Alternating twisted multilayer graphene presents a compelling multiband system, with a coexistence of Dirac bands and flat bands, for exploring superconductivity. However, the roles of flat bands and dispersive bands played in determining the superconductivity remain elusive. Here, we focus on the alternating twisted quadralayer graphene to reveal unconventional superconducting behaviors by systematically quantifying individual contributions for both the dispersive bands and the flat bands. The superconductivity is robust, with a strong electrical field tunability, a maximum BKT transition temperature of 1.6 K, and high critical magnetic fields beyond the Pauli limit. By analyzing the Landau fan diagram at zero electric displacement fields, we disentangle Dirac bands and flat bands, revealing a Coulomb interaction-induced band broadening effect. We further quantify the electric-field-dependent evolution of the critical temperature and coherence length, and estimate the flat-band Fermi velocity and superfluid stiffness via critical current measurements. Our results demonstrate an electric field–tunable coupling strength within the superconducting phase, revealing unconventional properties with vanishing Fermi velocity and large superfluid stiffness. These phenomena, attributed to substantial quantum metric contributions mediated by Dirac band hybridization, offer new insights into the mechanisms underlying unconventional flatband superconductivity in moiré systems.

**INTRODUCTION**

Superconductivity is characterized by flows of condensed electron pairs with zero dissipation. To understand the superconductivity, especially the multiband superconductors with high temperature superconductivity, is of fundamental interests and great promise for modern science and technology[1,2]. Recently, moiré superlattices composed of twisted multiple atomic thin layers give birth to topological flat band systems[3], where strong correlation effects enable the observation of superconductivity[4–12] along with correlated insulators[13–19], ferromagnetism[20–22], fractional[23–25] or integer[26–28] quantum anomalous Hall effect, etc. In twisted bilayer graphene (TBG), the flat band parts are situated in moiré AA regions at a low energy, while the dispersive band parts are distributed among moiré regions at a higher energy in the band structure[29]; superconductivity is observed in a specific magic angle[4]. A more intriguing example is the alternating twisted multilayer graphene (ATMG) systems[7,8,10–12,30,31], a typical multiband system hosting dispersive and flat bands[32]. The number of dispersive bands is $N - 2$, where $N$ is the number of graphene layers. ATMG have opened new avenues to investigate robust unconventional superconductivity, hosting a widely electric field-tunable superconducting phase diagram with an enhanced critical temperature and a large in-plane magnetic field that exceeds the Pauli limit[33]. The hybridization between dispersive and flat bands, tunable by electrical displacement fields, triggers significant changes in superconducting coupling strength, enabling the investigation of Bardeen-Cooper-Schrieffer (BCS) to Bose-Einstein condensate (BEC) crossover[7,12]. Despite these efforts, the nature of superconductivity in ATMG remains elusive, particularly regarding the individual roles of dispersive bands and flat bands in the electric-field tunable superconducting behaviors. Resolving these uncertainties would shed light on the mechanism of superconductivity in twisted moiré systems and other systems as well.

Here, we focus on the superconductivity in alternating twisted quadralayer graphene (ATQG) with a twisted angle of 1.58°. As a member of ATMG family, ATQG is an ideal candidate to investigate the superconductivity in multi-bands systems, containing two dispersive bands and one flat band. We elucidate carrier contributions from flat and dispersive bands by analyzing Landau fan diagrams, and we quantify the Fermi velocity of the flat bands from critical current measurements outside the superconducting region and the superfluid stiffness in the superconducting region. These quantities, together with the electric field-tunable superconducting behaviors, demonstrate the unconventional nature of the superconductivity in ATQG, where rich interplays among Coulomb interaction, Dirac bands mediated quantum metric, and flat band superconductivity are highlighted.

**Robust Superconductivity in ATQG System**

The ATQG sample is fabricated by using the 'cut and stack' method[34]. The four monolayer pieces, cutting from one large flake of monolayer graphene, are alternately stacked with one relative twisted angle $θ$ (Fig. 1a), resulting in one single moiré unit cell with the size defined by $θ$. In ATQG, two Dirac-type bands, located at the corner of moiré Brillouin zone $K_s$ and $K_s$', respectively, coexist with the flat band[32,35], as shown in Fig. 1b. These Dirac bands are independent from the flat band at interlayer potential energy $U = 0$, while they hybridize with the flat band at $U \neq 0$ with a gap opening at CNP (Fig. 1c). Fig. 1d exhibits a typical longitudinal resistance ($R_{xx}$) map as a function of the total filling factor $v_{tot}$ and the displacement field $D$ at a base temperature of $T = 30$ mK. Here, we define $v_{tot}$ = $v_{flat} + v_{Dirac} = 4n/n_0$, where $v_{flat}$ and $v_{Dirac}$ correspond to the filling factor of the flat band and Dirac bands, respectively. The full filling carrier density $n_0$ are extracted from the full filling resistance peak at $U \neq 0$ where hybridization eliminates the discrepancy between $v_{tot}$ and $v_{flat}$. It shows $n_0 = 5.78 \times 10^{12}$ /cm$^2$, corresponding to a twisted angle of 1.58°, where the bandwidth of flat band is ~10 meV in continuum model calculations (Fig. 1b).

We observe robust superconductivity in a wide range of the phase diagram in Fig. 1d, indicated by the two large blue-colored regions with zero resistance spanning from $v_{tot} \sim \pm 2$ to $\sim \pm 4$ and within a finite $|D| < \sim 0.5$ V/nm. As shown in Fig. 1e and 1f, the resistance in these regions quickly drops to zero when the temperature is lower than 1 K. At a fixed $v_{tot} = 2.31$, the temperature-dependent resistance exhibits a typical non-linear superconducting transition with $T_c$ of ~1.2 K at $D = 0$ V/nm and of ~2 K at $D = 0.26$ V/nm. Notably, as shown in Fig. 1g, the temperature-dependent voltage-current ($V$-$I$) curves show a clear Berezinskii–Kosterlitz–Thouless (BKT) transition at the critical temperature of $T_{BKT} = 1.1$ K, where $V$ is proportional to $I^3$, demonstrating the 2D superconducting nature in ATQG[36,37]. Furthermore, evidences of phase coherence are revealed in the magnetic field response of differential resistance d$V$/d$I$, showing a weak interference pattern and a strong suppression at critical magnetic field $B_\perp \sim 0.1$ T (inset of Fig. 1e). All the above-mentioned observations indicate these zero resistance states in the phase diagram are robust superconducting states.

**Coexistence of Dispersive and Flat Bands in ATQG**

Compared with twisted bilayer graphene (TBG), the involvement of dispersive bands extends the regime of superconductivity and simultaneously introduces complexity into the phase diagram[7,8,10–12]. Before investigating the mechanism of superconductivity in ATQG, it is crucial to disentangle the transport behaviors for the two types of bands and elucidate the respective contributions to carrier densities, which can be probed from the Landau fan diagram (Supplementary Note 4). At a fixed Fermi level, both Dirac bands and flat bands tend to form LLs due to Landau quantization; while the small Fermi surfaces in the former lead to LLs at small magnetic fields, the much larger Fermi surfaces in flat bands result in LLs at much larger magnetic fields. The total Chern number is the sum of that in flat bands and that in Dirac bands. In particular, above certain high magnetic fields, the total Chern number is the sum of the Chern numbers of the flat band LLs and that of the Dirac bands zero-LLs.

Fig. 2a-c show Landau fan diagrams at $D = 0$. On the conduction band side ($v_{tot} > 0$), three sets of Landau level (LL) sequences, each starting from the LL filling factor (Chern number) $v_{LL} = +4$, emerge from $v_{tot} \sim 2$, ~3.3, and 4, respectively. The extraction details of Chern number and the band filling factor are presented in Supplementary Note 3. The $v_{LL} = +4$ LLs, clearly visible in the Hall resistance map of Fig. 2b (also see Supplementary Fig. 2), arise from the quantization of two Dirac bands within a topologically trivial gap in the flat band. Additionally, two sequences of higher LLs with indexes of +12, +20, and +28 are observed emanating from $v_{tot} = 3$ and 3.3 (see details in Supplementary Fig. 3), which is consistent with a doubling of the LL index structure expected from a single Dirac band (i.e.,

+6, +10, +14). At $v_{tot} > 4$, another LL sequence with $v_{LL}$= +4, +12, +20, +28, +36 appears, suggesting that the exclusive contribution from Dirac bands to the LL spectrum in this regime. The emergence of LLs from nonzero filling factors suggests interaction-induced energy gaps[38] at integer fillings 2 and 3, as well as at the fractional filling 10/3 of the flat band. We attribute the ground state at fractional filling 10/3 to a charge density wave state, analogous to states previously reported in alternating twisted trilayer graphene[30,39].

We quantitatively extract the chemical potential and carrier density of Dirac bands[7,30] based on the trajectories of the $R_{xx}$ maxima (Supplementary Fig. 4) where the chemical potential lies within the $N$th LL of Dirac bands. The chemical potential difference between the $N$th LL ($N_{LL} > 0$ for the conduction band) and the zeroth LL is expressed as $\mu_N - \mu_0 = v_F\sqrt{2e\hbar N_{LL} B}$, where $v_F$ is the Fermi velocity of Dirac bands and $\hbar$ is the reduced Planck constant. According to the continuum model, $v_F$ is renormalized to ~$6.1 \times 10^5$ m/s. The carrier density of Dirac bands is given by $n_{Dirac} = 2N_{LL} \times 4eB/h$. Using the extracted chemical potential values from the $N_{LL} = 1$ and $N_{LL} = 2$ trajectories (Fig. 2d), we estimate the bandwidth of the flat band to be ~50 meV, significantly larger than the ~10 meV predicted by the continuum model. This large discrepancy suggests the flat band is non-rigid, where strong electron-electron interactions induce a band broadening effect through the formation of many-body energy gaps[40]. In Fig. 2e, the carrier density contributed by Dirac bands is ~ $0.96 \times 10^{12}$ /cm$^2$ at $v_{tot} = 4$, one order larger than the single particle simulation prediction. This again supports the presence of substantial band broadening due to interactions.

Furthermore, both the chemical potential and carrier density remain nearly constant from $v_{tot} \sim 1.35$ to $v_{tot} \sim 2.04$, indicating the flat band is highly compressible within this doping range. Subsequently, both the chemical potential and carrier density increase rapidly for $v_{tot} > \sim 2.04$ ($v_{flat} \sim 1.76$), suggesting the onset of an interaction-driven cascade transition, where the internal symmetry (spin/valley) is broken and the symmetry-broken sub-flat-band (isospin polarized) starts to fill again, along with Dirac bands. To account for interaction effects[41], we introduce an isospin polarization energy near half filling (Fig. 2f). At $v_{flat} = 2$, the Hartree interaction (Coulomb repulsion energy) raises the flat band relative to the Dirac bands, leading to an increased carrier occupation in the latter. Meanwhile, the Fock interaction (exchange energy) splits the isospin sub-bands with an energy shift of $2\Delta_{HF} = 40$ meV. This picture effectively reproduces the substantial carrier occupations of the Dirac bands and the isospin-polarized state at $v_{flat} = 2$.

The distinction between $v_{flat}$ and $v_{tot}$ is further supported by Hall resistance measurements. As shown in Supplementary Fig. 6a, two regions of maximum Hall resistance appear near $v_{tot} = \pm 2$ at low displacement fields, marked by yellow dashed lines. These maxima indicate that the Hall carrier density, given by $n_H = B/(eR_{xy})$, resets to almost zero, consistent with a cascade transition of the flat band near half filling. Assuming minimal complexity, we regard the Hall carrier density as an estimate of the real carrier density. As shown in Fig. 2g, $n_H$ almost equals to $n_{Dirac}$ at $v_{tot} = 2.3$ ($v_{flat} = 1.94$), implying that the flat band contributes negligible free carriers near half filling due to the symmetry-breaking cascade transition. Beyond full filling ($v_{tot} > 4$), the Hall carrier density follows a linear relation with the doping level, $n_H = (v_{tot} - v_0) n_0/4$, suggesting the carrier density is exclusively contributed by Dirac bands in this regime.

**Electrical Field Tunable Coupling Strength of Superconductivity**

To explore the superconductivity in this multi-band system, we apply a displacement field $D$ to introduce band hybridization and investigate its effect on the coherent length and critical temperature. Fig. 3a exhibits the superconducting phase diagram as a function of $D$ at $v_{tot}$ = 2.31. By applying $D$ from 0 to 0.25 V/nm, the superconductivity is enhanced with $T_c$ rising from 1.2 to 2 K; further increasing $D$, $T_c$ decreases and eventually vanishes at $D > \sim 0.4$ V/nm. To quantitatively estimate the coupling strength of Cooper pairs in the superconducting phase, we compare the superconducting coherence length and inter-particle distance $d$. The Ginzburg – Landau (G-L) coherence length $\xi$ is extracted from the critical magnetic field, i.e., $\xi = \sqrt{\Phi_0/(2\pi B_{c0})}$, where $\Phi_0 = h/(2e)$ is the superconducting flux quantum and $B_{c0}$ is the critical magnetic field extrapolated to the zero-temperature limit. The measured G-L coherence length $\xi$ shows an inverse trend of $T_c$, reaching a minimum $\xi$ of ~30 nm at $D = 0.25$ V/nm (Fig. 3b). The inter-particle distance $d$ is approximated as inversely proportional to the Fermi vector $k_F$, i.e., $1/d \sim k_F = \sqrt{4\pi n^*/g}$, where $n^*$ is the pairing carrier density and $g = 2$ is the band degeneracy by considering the symmetry breaking at half filling. We define $n^*$ or equivalently $v^*$ as $v^* = |v_{flat} - 2|$, where $v^*$ is measured relative to the half filling of the flat band. Fig. 3c shows the product of $\xi$ and $k_F$, which characterizes the ratio of the coherence length to the inter-particle distance and thus serves as a metric for the coupling strength. $\xi \times k_F$ remains below 6 across the entire $D$ range and reaches a minimum of 2 at $D = 0.25$ V/nm, suggesting the superconductivity is modulated within a narrow regime close to the BEC limit[42].

The strongly coupled superconductivity modulated by $D$ exhibits unconventional behaviors. As shown in Fig. 3d, $n_H$ extracted at $B_\perp = 0.5$ T (where superconductivity is fully suppressed), remains constant for $|D| < 0.3$ V/nm and then diverges rapidly. This divergence in $n_H$ suggests an enhanced DOS near a van Hove singularity (VHS). However, $T_c$ decreases and the coupling strength weakens for $|D| > 0.3$ V/nm. This observation is contrary to conventional BCS expectations. Furthermore, the $dV/dI$ curves in Fig. 3e reveal a higher critical current at finite $D$ compare to $D = 0$, suggesting a stronger coupling superconductivity approaching the BEC regime. The negative temperature-dependent behavior of $dV/dI$ (Fig. 3f) in the normal state above critical current deviates from the conventional metal behavior. The strong correlation between electrons potentially leads to a pseudo energy gap of pre-paired electrons[12]. However, further experiments and theoretical investigations are needed to fully understand this phenomenon.

**Critical currents and superfluid stiffness**

We further quantify the superconducting properties through critical current measurements. Doping dependent $dV/dI$ maps at $D = 0.26$ V/nm reveal two distinct boundaries characterized by different critical currents (Fig. 4a-b). The lower boundary $j_s$ vanishes with increasing perpendicular magnetic field, indicating its association with the superconducting critical current. In contrast, the upper boundary $j_n$ remains evident even at higher magnetic fields. This behavior suggests that $j_n$ originates from the critical behavior of a correlated gap at half filling, arising from interband transitions via the Zener-Klein tunnelling[43–46] at high current densities. We estimate the Fermi velocity of the flatband from the drift velocity ($v_n$) associated with the higher critical current density using the formula $v_n = \frac{1}{e}\frac{dj_n}{dn}$. The resulting value $v_n \approx 1700$ m/s suggests an extremely flat band dispersion, comparable to that observed in magic angle TBG[45]. With the drift velocity, the superconducting gap is estimated as $\Delta \approx \hbar v_n/\xi$. Using $\xi = 30$ nm and $T_c = 2$ K at $D = 0.26$ V/nm, we obtain $\Delta/k_B T_c \approx 0.2$, one order lower than the BCS prediction of 1.75. In addition, we calculate the superfluid stiffness at zero temperature limit using the critical current density $j_s$ in Fig. 4c, based on the formula of $D_s = \frac{2\pi j_s \xi}{\Phi_0}$ in flat band condition[45] (see Supplementary Note 8). As shown in Fig. 4d, $D_s$ reaches the maxima of $\sim 4 \times 10^7$ /H around $D = 0.26$ V/nm, one order larger than the conventional contribution of $D_s = \frac{e^2 n^*}{m^*}$ (the red dashed line in Fig. 4d), where $m^* = \hbar k_F / v_n$ is the effective mass. As $D$ drives the system close to VHS, $D_s$ decreases rapidly and approaches the level of conventional contribution.

The enhanced coupling strength and superfluid stiffness tuned by $D$ can be attributed to the increasing contribution from the quantum metric. In flat band superconductivity, quantum metric contributions are comparable to or even larger than conventional contributions[47–49], and can be further amplified by the band touching in multiband systems[50]. According to the continuum model, increasing $D$ causes the band hybridization, which enhances the quantum metric near the flat band minimum. Due to the correlation effect, the Fermi energy reaches the flat band minimum near $v = 2$, where the quantum metric contribution is most significant. As shown in Fig. 4e and 4f, we calculate quantum metric contributions at $v_{flat} = 2$. The interlayer potential ($U$) hybridizes the flat band and Dirac bands at $K_s$ and $K_s$' points, and simultaneously, it generates hot spots of quantum metrics. As a result, the quantum metric can be enhanced at finite displacement fields, owing to the band hybridization. To make a comparison with experiment qualitatively, we further calculate the second order correlation function (details in Supplementary Note 14), which determines the coherence length. As shown in Fig. 4g, it reaches a maximum at a finite $U$, consistent with the enhanced superfluid stiffness in Fig. 4d. The agreement between experiments and the calculations establishes an intimate connection between quantum metric effect and flat band superconductivity, and it further highlights an important role played by the Dirac band in stabilizing multiband superconductivity via generating quantum metric hot spots.

In addition, as shown in Supplementary Fig. 10, the ratio between onset temperature $T^*$ and $T_{BKT}$ increases from 1.45 to 2.19 with $D$, supporting the emergence of strong coupling induced by quantum metric[51] at large $D$. Notably, measurements of the superfluid stiffness provide direct insight into the nature of the superconducting order parameter[52,53]. Recent kinetic inductance measurements also reveal possible nodal superconductivity in ATMG[54] and indispensable contribution of quantum metric to superfluid stiffness in TBG[55].

**CONCLUSIONS**

In conclusion, we investigate unconventional superconductivity in a field-tunable multi-band system and clarify the distinct roles of flat and dispersive (Dirac) bands. Remarkably, the strong coupling superconductivity emerges near the half filling of flat band and can be enhanced by applying $D$.

Moreover, we reveal the unconventional superconductivity with vanishing Fermi velocity and large superfluid stiffness, which goes beyond the conventional BCS prediction and could be ascribed to significant quantum metric contribution driven by $D$ when the Dirac bands touch flat band. Our findings provide new insights into the nature of superconductivity and the BCS-BEC crossover in multiband flat-band systems, with broader implications for engineered quantum materials.

**Acknowledgments**

We thank Fengcheng Wu, Zhida Song, Kun Jiang, Shiliang Li, and Kui Jin for useful discussions. We acknowledge support from the National Key Research and Development Program (Grant No. 2020YFA0309600), the Natural Science Foundation of China (NSFC, Grant Nos. 12074413 and 61888102), the Strategic Priority Research Program of CAS (Grant No. XDB33000000), and the Key-Area Research and Development Program of Guangdong Province (Grant No. 2020B0101340001). K.W. and T.T. acknowledge support from the Elemental Strategy Initiative conducted by the MEXT, Japan (Grant No. JPMXP0112101001), JSPS KAKENHI (Grant Nos. 19H05790, 20H00354 and 21H05233) and A3 Foresight by JSPS.

**Author contributions**

W.Y. and G.Z. supervised the project. W.Y. and L.L. designed the experiments. L.L. and Y.H. fabricated the devices and performed the magneto-transport measurement. L.L. performed the band structure calculations. C.Z. and K.L developed the theory for the quantum metric calculations. K.W. and T.T. provided hexagonal boron nitride crystals. L.L., C.Z., K.L., G.Z., and W.Y. analyzed the data. L.L. and W.Y. wrote the paper with the input from all the authors.

**Data availability**

The data that support the findings of this study are available from the corresponding authors upon reasonable request.

**Competing interests**

The authors declare no competing interests.

**Additional information**

Supplementary information is provided online

**Figure & Figure captions**

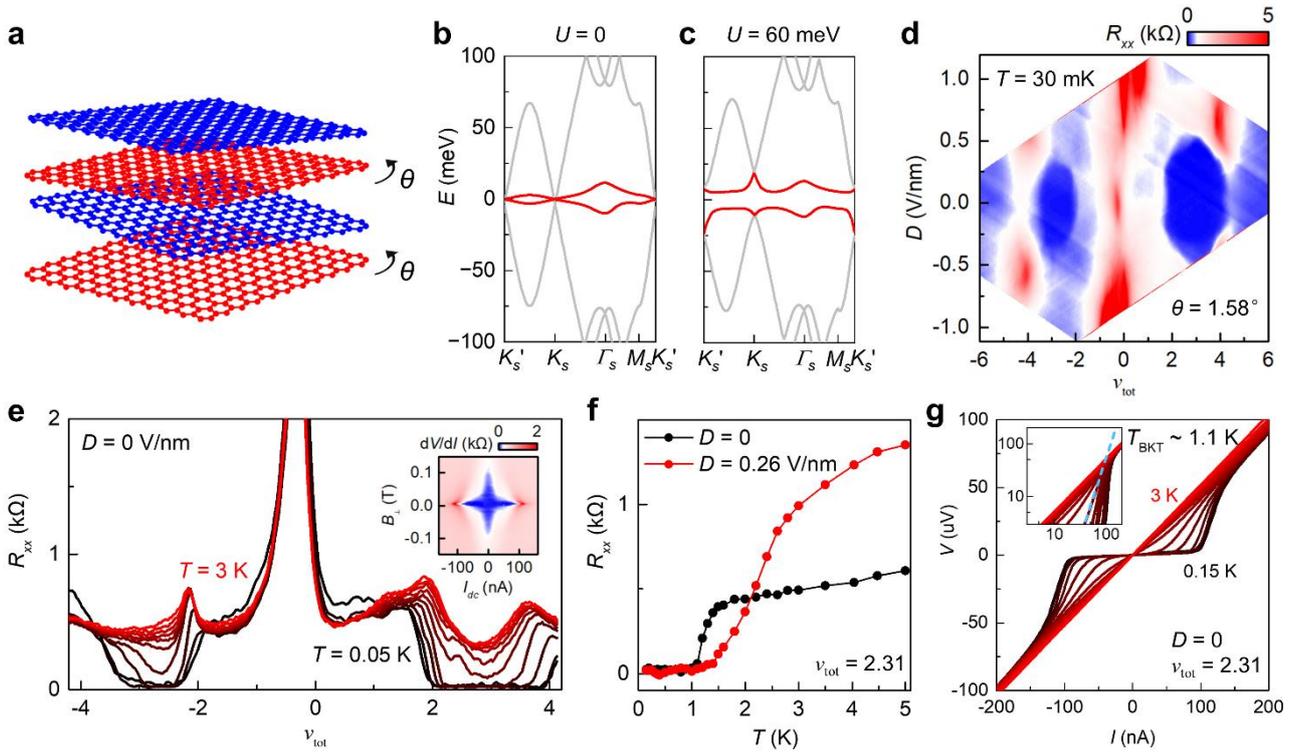

FIG. 1. Superconductivity in alternating twisted quadralayer graphene. **a**, Schematics of the stacking order of ATQG. **b**, **c**, Band structures calculated from the continuum model at $U = 0$ and $U = 60$ meV. **d**, $R_{xx}$ as a function of $v_{tot}$ and $D$ at $T = 30$ mK and $B = 0$ T. **e**, $R_{xx}$ versus $v_{tot}$ from $T = 0.05$ K to 3 K at $D = 0$. Inset, $dV/dI$ as a function of $I_{dc}$ and $B$ at $v_{tot} = 2.31$ and $D = 0$. **f**, $R_{xx}$ vesus $T$ at $D = 0$ and 0.26 V/nm. **g**, $V$ versus $I$ from $T = 0.15$ K to 3 K. Inset, zoom-in figure in log-log coordinate. The blue dash line is the nonlinear fitting of $V \sim I^3$.

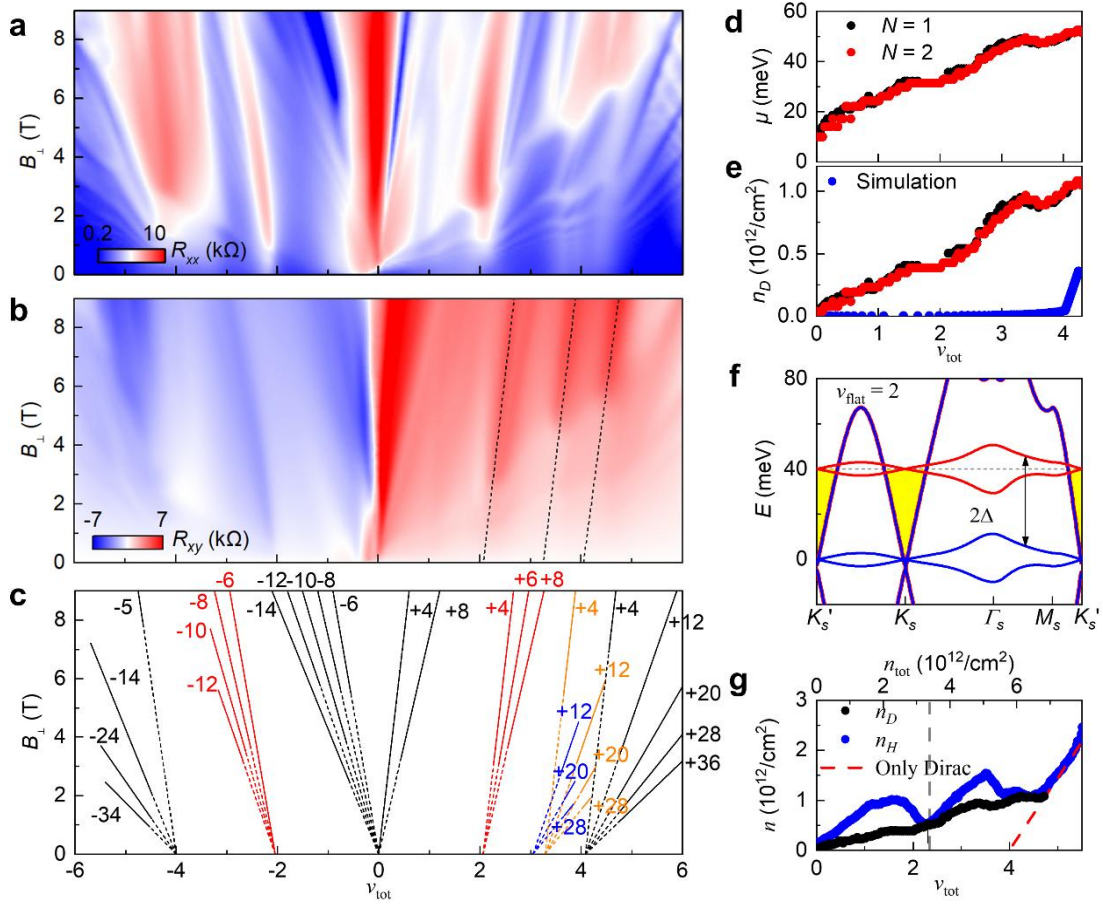

FIG. 2. Coexistence of flat and Dirac bands. **a-c**, Landau fan diagrams at $D = 0$ and $T = 1.7$ K. **d**, $\mu$ versus $v_{tot}$ of $N = 1$ and 2 LL from the Dirac band. **e**, $n_{Dirac}$ verus $v_{tot}$ of $N = 1$ and 2 LL from the Dirac band. The blue dot line is simulated carrier densities of Dirac bands. **f**, Band structure of the ATQG near $v_{flat} = 2$, where an isospin polarization is established due to the correlation effect. The dashed line indicates the chemical potential at 40 meV. The yellow shaded region marks the filled Dirac bands. $\Delta_{HF} = 20$ meV accounts for the flavor polarization at $v_{flat} = 2$. **g**, Carrier densities versus $v_{tot}$. The gray dashed line corresponds to $v_{flat} = 2$. The red dashed line corresponds to the fitting curve of $n = (v_{tot} - v_0) n_0/4$.

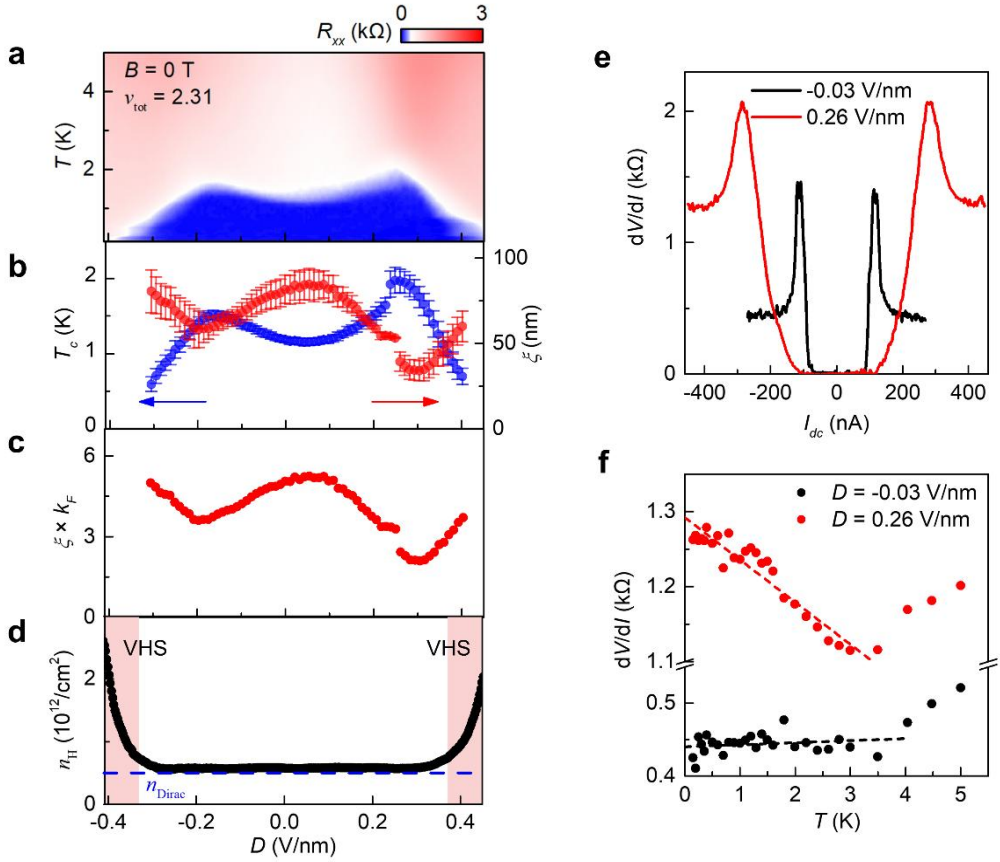

FIG. 3. *D* tunable coupling strength of superconductivity. **a**, $R_{xx}$ as a function of *D* and *T* at $\nu_{tot}$ = 2.31. **b**, $T_c$ and $\xi$ versus *D*. **c**, $\xi \times k_F$ versus *D*. **d**, $n_H$ versus *D*. The blue dashed line corresponds to the carrier density of Dirac bands. The pink regions are in the vicinity of VHS. **e**, d*V*/d*I* versus $I_{dc}$ at different *D*. **f**, d*V*/d*I* versus *T*. The data is measured at $I_{dc}$ = 265 (420) nA and *D* = -0.03 (0.26) V/nm.

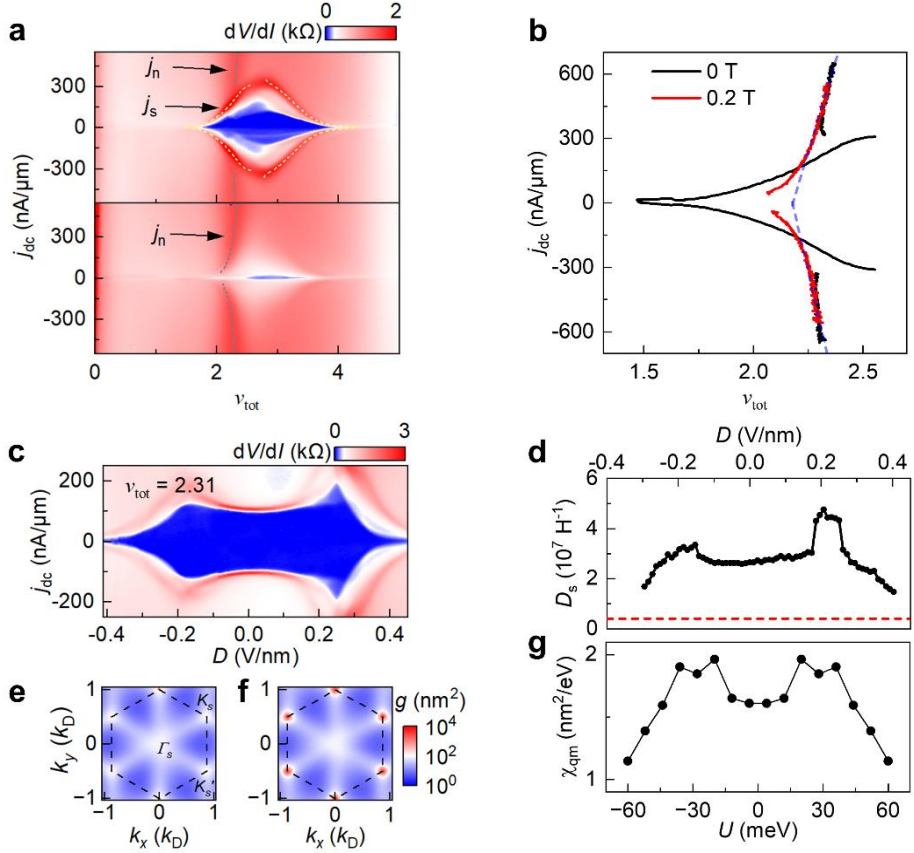

FIG. 4. Critical currents and superfluid stiffness. **a**, d$V$/d$I$ as a function of $v_{tot}$ and $I_{dc}$ at $D$ = 0.26 V/nm. Top, $B$ = 0 T. Bottom, $B_\perp$ = 0.2 T. **b**, Extracted critical currents for superconductivity and correlated insulator. Blue dashed lines are linear fitting curves. **c**, d$V$/d$I$ as a function of $D$ and $I_{dc}$ at $v_{tot}$ = 2.31. **d**, calculated superfluid stiffness versus $D$. The red dashed line corresponds to conventional contribution without considering quantum metric. **e-f**, Momentum space distribution of the quantum metric. $g$ is the trace of the quantum metric tensor. $k_D$ is the momentum difference between $K_s$ and $K_s$' valleys in the moiré Brillouin zone. **g**, The correlation function versus $U$.